\documentclass[reprint,
superscriptaddress,
 amsmath,amssymb,
 aps,
prb,
]{revtex4-2}

\usepackage[utf8]{inputenc}
\usepackage{mathtools,csquotes}
\usepackage{amssymb}
\usepackage{bm}
\usepackage{multirow}
\usepackage{makecell}
\usepackage{nicematrix}
\usepackage{hyperref}

\newcommand{\dd}{\mathrm{d}}
\newcommand{\ii}{\mathrm{i}}

\newcommand{\mc}[1]{\mathcal{#1}}
\newcommand{\bb}[1]{\mathbb{#1}}
\newcommand{\hb}[1]{\Hat{\mathbf{#1}}}

\begin{document}

\preprint{APS/123-QED}

\title{Nonabelian elastic metamaterials using holonomies acquired by crossing degeneracies}

\author{Mohit Kumar}
\affiliation{Ray W. Herrick Laboratories, School of Mechanical Engineering, Purdue University, West Lafayette, Indiana 47907, USA}

\author{Ralph M. Kaufmann}
\email{rkaufman@purdue.edu}
\affiliation{Department of Mathematics, Purdue University, West Lafayette, Indiana 47907, USA}
\affiliation{Department of Physics and Astronomy, Purdue University, West Lafayette, Indiana 47907, USA}
\affiliation{PQSEI, Purdue University, West Lafayette, Indiana 47907, USA}

\author{Fabio Semperlotti}%
\email{fsemperl@purdue.edu}
\affiliation{Ray W. Herrick Laboratories, School of Mechanical Engineering, Purdue University, West Lafayette, Indiana 47907, USA}

\date{\today}

\begin{abstract}
Embedding nonabelian features into elastic metamaterials promises remarkable opportunities for wave control in many practical applications such as surface acoustic wave devices, mode multiplexers, and on-material computation. Nevertheless, current realizations are limited to arrangements of coupled resonators with fine-tuned interactions, limiting their applicability to continuous media. This theoretical and numerical study introduces a design principle for continuous nonabelian elastic metamaterial waveguides. The basic configuration consists of a composite waveguide made of multiple cylindrical waveguides coupled by spatially varying elements. These elements are engineered to follow geometrically-controlled parameter variations that cross selected degeneracies and produce a targeted nonabelian holonomy. The strategy based on crossing degeneracies fundamentally differs from abelian geometric phases, where parameters avoid and encircle degeneracies, or nonabelian Wilczek-Zee phases, where parameters are fine-tuned to maintain degeneracies throughout their cycle. The resulting holonomy transfers an input longitudinal excitation in one rod to an output response in another rod. When two such waveguides are concatenated, their ordering dictates the output response, thereby revealing the emergence of nonabelian dynamics. The nonabelian behavior persists across a broad range of frequencies and under perturbations to the geometry of coupling elements or cylinder diameters. These results establish a robust, effective, and practical route to leverage nonabelian physics in elastic metamaterials.
\end{abstract}

\maketitle

\section{Introduction}
Nonabelian metamaterials are emerging as a powerful platform to control and manipulate the dynamics of classical waveguide systems, providing pathways towards mode-multiplexed devices, robust mode converters, and on-material computation~\cite{yang_non-abelian_2024,fruchart_dualities_2020,zanardi_holonomic_1999,zhang_geometric_2023,nayak_non-abelian_2008}. These metamaterials are designed to embed noncommutative features into synthetic gauge fields~\cite{aidelsburger_artificial_2018,chen_non-abelian_2019,cheng_non-abelian_2025,wu_non-abelian_2022,yan_non-abelian_2023}, topological invariants of the band structure~\cite{sato_non-abelian_2009,wu_non-abelian_2019,jiang_experimental_2021,yang_homotopy_2024}, or propagation of wave modes~\cite{fruchart_dualities_2020,iadecola_non-abelian_2016,noh_braiding_2020,chen_classical_2022,you_observation_2022,sun_non-abelian_2022}, which ultimately lead to a path-dependent evolution of physical or abstract observables. Despite significant theoretical and experimental progress in photonic, acoustic, and discrete mechanical systems, realizing nonabelian physics in continuous elastic systems remains challenging, in part due to the limitations of coupled-resonator descriptions. Achieving nonabelian behavior in elastic metamaterials would unlock an untapped design space to control mechanical waves, with implications for smart structures and surface acoustic wave devices.

To design a nonabelian elastic waveguide, existing acoustic~\cite{you_observation_2022,chen_classical_2022} and photonic~\cite{yang_non-abelian_2024,brosco_non-abelian_2021,kremer_optimal_2019,zhang_non-abelian_2022,shan_non-abelian_2025,chen_high-dimensional_2025,song_shortcuts_2025} implementations point to the concept of holonomy, which is the change in an elastic guided mode after a slow, cyclic variation in the cross-sectional parameters of the waveguide~\cite{berry_quantal_1984,simon_holonomy_1983,wilczek_appearance_1984,cohen_geometric_2019,cisowski_colloquium_2022,deymier_sound_2017}. There are three strategies to acquire holonomies, classified by the parameter space loop traced by the cross-sectional variations (Figs.~\ref{fig:design-schematic}a-c). The strategy required here is the one that results in a robust and noncommuting holonomy. Robustness is needed to avoid fine-tuning or costly optimization in the design process. Noncommutativity ensures that the final response after parameter cycles depends on their ordering, giving rise to nonabelian dynamics.

The first strategy uses parameter variations that encircle degeneracies (Fig.~\ref{fig:design-schematic}a). The resulting holonomy is the well-known topological geometric phase, where an initial guided mode $\mathbf{W}$ reverses its sign and evolves to $-\mathbf{W}$~\cite{berry_quantal_1984,boulanger_observation_2012,snieder_seismic_2016,tromp_berry_1992}. This strategy is robust because small perturbations to the parameter evolution still encircle the degeneracy, preserving the holonomy. However, the holonomies are abelian because it essentially implements scalar multiplication.

The second strategy uses parameter variations during which degeneracies are maintained between three or more guided modes (Fig.~\ref{fig:design-schematic}b). In parameter space, the loop lies on a low-dimensional manifold supporting degeneracies. This results in the well-known Wilczek-Zee phase~\cite{wilczek_appearance_1984}, where an initial guided mode $\mathbf{W}$ transforms by a rotation matrix $\mc{R}$ to evolve a final guided mode $\mc{R} \mathbf{W}$. Since rotation matrices are noncommuting, the holonomy is nonabelian. However, the holonomy is extremely sensitive to small perturbations to the parameter variations because these perturbations no longer maintain the degeneracies. As a result, existing realizations are limited to acoustic, photonic, or discrete mechanical systems accurately modeled as coupled resonators~\cite{yang_non-abelian_2024,brosco_non-abelian_2021,kremer_optimal_2019,zhang_non-abelian_2022,shan_non-abelian_2025,chen_high-dimensional_2025,song_shortcuts_2025,chen_observation_2025,lu_experimental_2025,meng_non-abelian_2025,fruchart_dualities_2020,yang_non-abelian_2023,iadecola_non-abelian_2016,noh_braiding_2020,ezawa_braiding_2019,ezawa_non-abelian_2020,zou_experimental_2023,chen_topological_2025,barlas_topological_2020,wu_observation_2024,saunders_experimental_2025}, which is certainly invalid in continuous elastic waveguides. Thus, this strategy is unlikely to yield robust nonabelian  behavior in elastic waveguides. 

The third strategy uses parameter variations that cross degeneracies at discrete parameter values (Fig.~\ref{fig:design-schematic}c). This recently introduced strategy was implemented in acoustic~\cite{you_observation_2022} and photonic~\cite{sun_non-abelian_2022,sun_two-dimensional_2024,sun_reconfigurable_2025} realizations of a nonabelian Thouless pump. The resulting holonomies are nonabelian and appear to be robust against perturbations, highlighting its suitability to create a nonabelian elastic waveguide. However, the theoretical background of such nonabelian holonomies, the design principles to acquire them, and their elastic realizations have not been pursued.

This study leverages degeneracy-crossing in parameter space to formulate a rigorous design principle for nonabelian elastic waveguides. The resulting composite waveguide comprises multiple cylindrical waveguides coupled by thin plates, and it achieves a targeted holonomy that transfers longitudinal excitations from one waveguide to another, as verified in numerical simulations. When two such waveguides are concatenated, the propagation of longitudinal excitations depends on the ordering of waveguides, signaling the emergence of nonabelian behavior. The nonabelian behavior is broadband and robust against perturbations to the waveguide geometry.

Furthermore, the study establishes the theoretical foundations of holonomies induced by degeneracy-crossing. The theoretical analysis shows that parameter cycles crossing degeneracies allow guided modes to be permuted in addition to changing their signs. The resulting set of achievable holonomies is characterized as the group of orientation-preserving signed permutation matrices and is connected to the well-known braid group, generalizing the findings of earlier studies that focused on permutations~\cite{you_observation_2022,sun_non-abelian_2022,sun_two-dimensional_2024}. The theory is extended to systems maintaining degeneracies during parameter variation, where eigenspaces of guided modes can be permuted, and to systems governed by Schr\"odinger-like equations. Additionally, in SI Appendix, Sec.~3, we show that a mathematical description of the holonomy is achieved by gluing connections on a fiber bundle.

\section{Approach to design elastic waveguides}
\subsection{Description of system's configuration}
\begin{figure*}
    \centering
    \includegraphics[width=\linewidth]{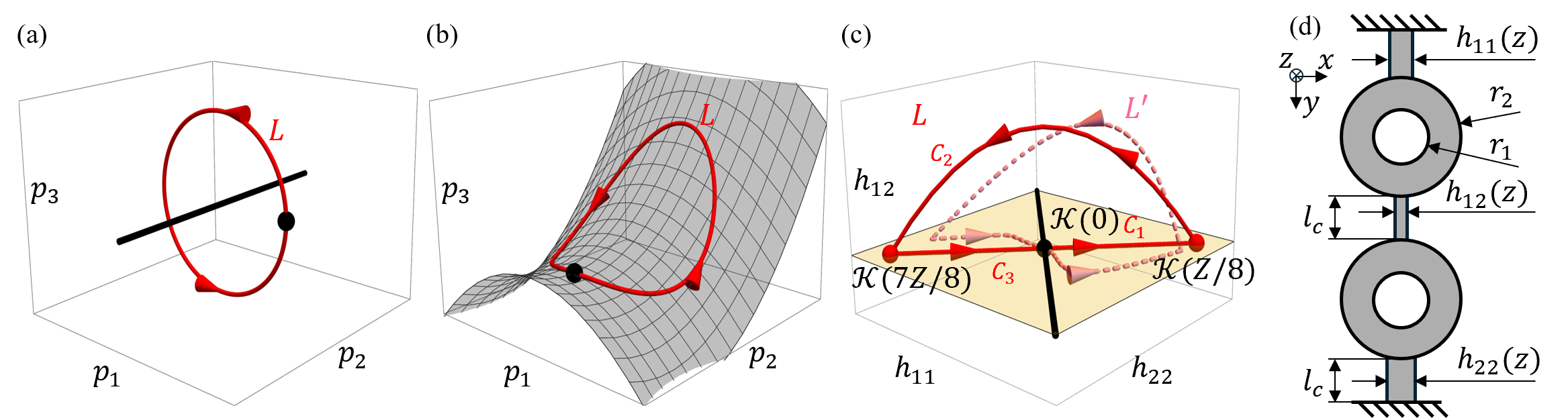}
    \caption{(a) Schematic illustration of parameter variations (red loop) that result in a topological geometric phase. The parameter space consists of three quantities $[p_1,p_2,p_3]$. The black solid line denotes parameter values supporting degeneracies in the system. The black dot denotes the start and end point of the loop. (b) Schematic illustration of parameter variations (red loop) that result in a Wilczek-Zee phase. The parameter space consists of three quantities $[p_1,p_2,p_3]$. The gray surface denotes parameter values supporting degeneracies in the system. The black dot denotes the start and end point of the loop. (c) Schematic illustration of parameter variations (red loop $L$) that cross a degeneracy and result in a holonomy. The 3D parameter space $[h_{11},h_{22},h_{12}]$ denotes the thickness of the coupling plates in a system of two coupled rods shown in (d). Each point represents a set of three values of the thickness of the coupling plates at a cross section. Loop $L$ parametrized by $z$ corresponds to a waveguide of length $Z$ with a cyclic variation in the cross section. It comprises line $C_1$, curve $C_2$, and line $C_3$. The points on loop $L$ at $z=0,Z/8,$ and $7Z/8$ also report the associated coupling matrix as $\mc{K}(z)$. Perturbing $L$ to $L'$ leaves the holonomy unchanged. The yellow plane at $h_{12}=\mathrm{const}=0$ denotes uncoupled rods. The black solid line marks twofold degeneracies, for which $h_{11}=h_{22}$ and $h_{12}=0$. (d) Cross section of the composite elastic waveguide made of $N=2$ cylindrical rods.}
    \label{fig:design-schematic}
\end{figure*}

Consider a composite elastic waveguide comprising $N$ identical rods having an annular cross section (i.e., cylindrical waveguides) of inner radius $r_1$ and outer radius $r_2$ (Fig.~\ref{fig:design-schematic}a). The rods have a modulus of elasticity $E$, density $\rho$, and bar velocity $c=\sqrt{E/\rho}$. Rods $i$ and $j$ ($i \neq j$) are coupled to each other with connecting plates of length $l_c$ and thickness $h_{ij}(z)$. By definition, $h_{ij}(z)=h_{ji}(z)$. The $i$th rod is also fixed to the ground by a plate of length $l_c$ and thickness $h_{ii}(z)$. In the following analyses, all plates have shear modulus $G_c$, and variable thickness along the $z$ direction. 

The longitudinal dynamics is modeled in the low-frequency regime; flexural and torsional dynamics are ignored as they do not interfere with the design and performance of the system. The longitudinal displacement of each rod $w_i(z,t)$ is approximately constant in its cross section, and the transverse displacements are neglected. The longitudinal displacements in rods $i$ and $j$ are assumed to interact exclusively through in-plane shear deformation of the connecting plate, which contributes an effective stiffness per unit length of $G_c h_{ij}(z)/l_c$. Similarly, the grounding plate contributes an effective stiffness per unit length of $G_c h_{ii}(z)/l_c$. The mass and dynamics of the plates are not incorporated into the model. With these modeling assumptions, the governing equation of motion is
\begin{equation}
\label{eqn:rod-full-eom}
    \rho A \Ddot{\mathbf{w}}+(G_c/l_c)\mathcal{K}(z) \mathbf{w} = E A \mathbf{w}''\;,
\end{equation}
where $\mathbf{w}(z,t)$ is the state vector of longitudinal displacements, overdots denote time derivatives, primes denote derivatives with respect to $z$, $A=\pi(r_2^2-r_1^2)$ is the cross-sectional area of each rod, and $\mathcal{K}(z)$ is the coupling matrix defined as
\begin{equation}
\label{eqn:coupling-matrix}
    \mc{K}_{ij}(z)=
    \begin{cases}
        \sum_{l=1}^n h_{il}(z)\;, & i = j\;,\\
        -h_{ij}(z)\;, & i \neq j\;,
    \end{cases}
\end{equation}
analogous to the stiffness matrix of a spring-mass system.  

At an operating frequency of $\omega$, the steady-state displacement response is harmonic, that is, $\mathbf{w}(z,t)=\hb{w}(z)e^{-\ii \omega t}$. Substituting this ansatz into Eq.~\ref{eqn:rod-full-eom} provides
\begin{equation}
\label{eqn:eom}
    \hb{w}''(z) + \mc{D}(z) \hb{w}(z)=0\;,
\end{equation}
where 
\begin{equation}
\label{eqn:dynamical-matrix}
    \mc{D}(z) = \frac{\omega^2}{c^2}\mc{I}_N - \frac{G_c}{EAl_c} \mc{K}(z)
\end{equation}
is termed the dynamical matrix and $\mc{I}_N$ is the square identity matrix of size $N$. Equation~\ref{eqn:eom} describes the propagation of longitudinal waves in the spatially-varying waveguide.

At each generic coordinate $z=z_0$, one can imagine to take a cross section (with a plane $xy$) of the composite waveguide and use the associated cross-sectional properties to build an equivalent (constant property) waveguide having thickness of the different connectors $h_{ij}(z)=h_{ij}(z_0)$. This equivalent uniform waveguide admits guided mode solutions $\mathbf{w}(z,t)=\mathbf{W}(z_0)e^{\ii k z - \ii \omega t}$, which simplifies Eq.~\ref{eqn:rod-full-eom} to the eigenvalue problem
\begin{equation}
\label{eqn:evp}
    \mc{K}(z_0) \mathbf{W}(z_0) = \frac{EAl_c}{G_c} \left( \frac{\omega^2}{c^2}- k(z_0)^2 \right) \mathbf{W}(z_0)\;,
\end{equation}
whose solution at a given frequency $\omega$ provides the corresponding guided mode displacements $\{ \mathbf{W}_1(z_0), \dots \mathbf{W}_N(z_0) \}$ and their wavenumbers $\{ k_1(z_0), \dots, k_N(z_0) \}$. The guided modes are collected into columns of a matrix $\mc{W}(z_0)$. As the guided modes are orthonormal, $\mc{W}(z_0)$ defines an orthonormal frame. By varying $z_0$, each cross section of the original spatially-varying waveguide is associated to guided modes and wavenumbers. This family of guided mode solutions, which is parameterized by the longitudinal coordinate, lays the foundation to understand the propagation of waves in spatially varying waveguides~\cite{bretherton_propagation_1968,tromp_berry_1992,kumar_role_2024,berry_quantal_1984,budden_phase_1976}.

\subsection{Design process to achieve target holonomies}

The waveguides in this study are designed to achieve a target holonomy identified by a signed permutation matrix $\mc{H}$. In other words, the waveguide will transform an input longitudinal excitation localized on rod $i$ to an output response localized on rod $j$ after propagating for a length $Z$. The wave field may also pick up a 180${}^\circ$ phase angle (that is, a sign-change) beyond the usual dynamical phase $\phi(Z)$, reminiscent of a topological geometric phase. In mathematical terms, the harmonic input excitation $\hb{w}(0)=\mathbf{e}_i$ will evolve to $\hb{w}(Z)=\mc{H}\mathbf{e}_i e^{\ii \phi(Z)}=\pm \mathbf{e}_j e^{\ii \phi(Z)}$, where $\mathbf{e}_i$ is the $i$th standard basis vector. Physically, $\mathbf{e}_i$ denotes a uniform unit longitudinal displacement in the $i$th rod; transverse displacements are always assumed zero in our discrete model because flexural and torsional dynamics minimally affect the design. 

\subsubsection{Design inputs and constraints}
During the design of the waveguide, the material constants $E$, $G_c$, and $\rho$, the radii of the cylinders $r_1$ and $r_2$, the length of the coupling plates $l_c$, and the maximum operating frequency $\omega_\mathrm{max}$ are assumed to be fixed quantities (see SI Appendix, Sec.~5A for numerical values). Then, the target holonomy matrix is specified as a product of matrices of size $N$:
\begin{equation}
\label{eqn:givens-holonomy}
    \mc{H}=\mc{G}(N-1,N,\theta_{N-1,N})\dots\mc{G}(1,2,\theta_{12})\;,
\end{equation}
where each matrix $\mc{G}(i,j,\theta_{ij})$ indicates a two-dimensional rotation that rotates the plane spanned by $\mathbf{e}_i$ and $\mathbf{e}_j$ by an angle $\theta_{ij}$ which equals $0$ or $\pi/2$. The generic matrix $\mc{G}(i,j,\theta_{ij})$ is known as a Givens rotation~\cite{golub_matrix_2013} and its formal description is presented in Eq.~\ref{eqn:givens-matrix} in the Materials and Methods section. This choice of the holonomy matrix ensures that the ensuing thickness of the coupling plates are nonnegative (positive or zero), as explained in SI Appendix, Sec.~4B. Using this target holonomy as the input, the following design process provides the variation in the thickness of coupling plates $h_{ij}(z)$ and the required length of the waveguide $Z$.

The design space is restricted to thin coupling plates, $h_{ij}(z)<h_\mathrm{max}=0.1l_c$, and low operating frequencies, $r_2 \omega_\mathrm{max}/c < 1.25$. In this regime the discrete model (Eq.~\ref{eqn:eom}) satisfactorily approximates the guided modes of the continuous waveguide obtained from finite element simulations on the commercial software COMSOL Multiphysics (SI Appendix, Sec.~5E). 

\subsubsection{Variation in plate thicknesses}
\begin{figure}
    \centering
    \includegraphics[width=\linewidth]{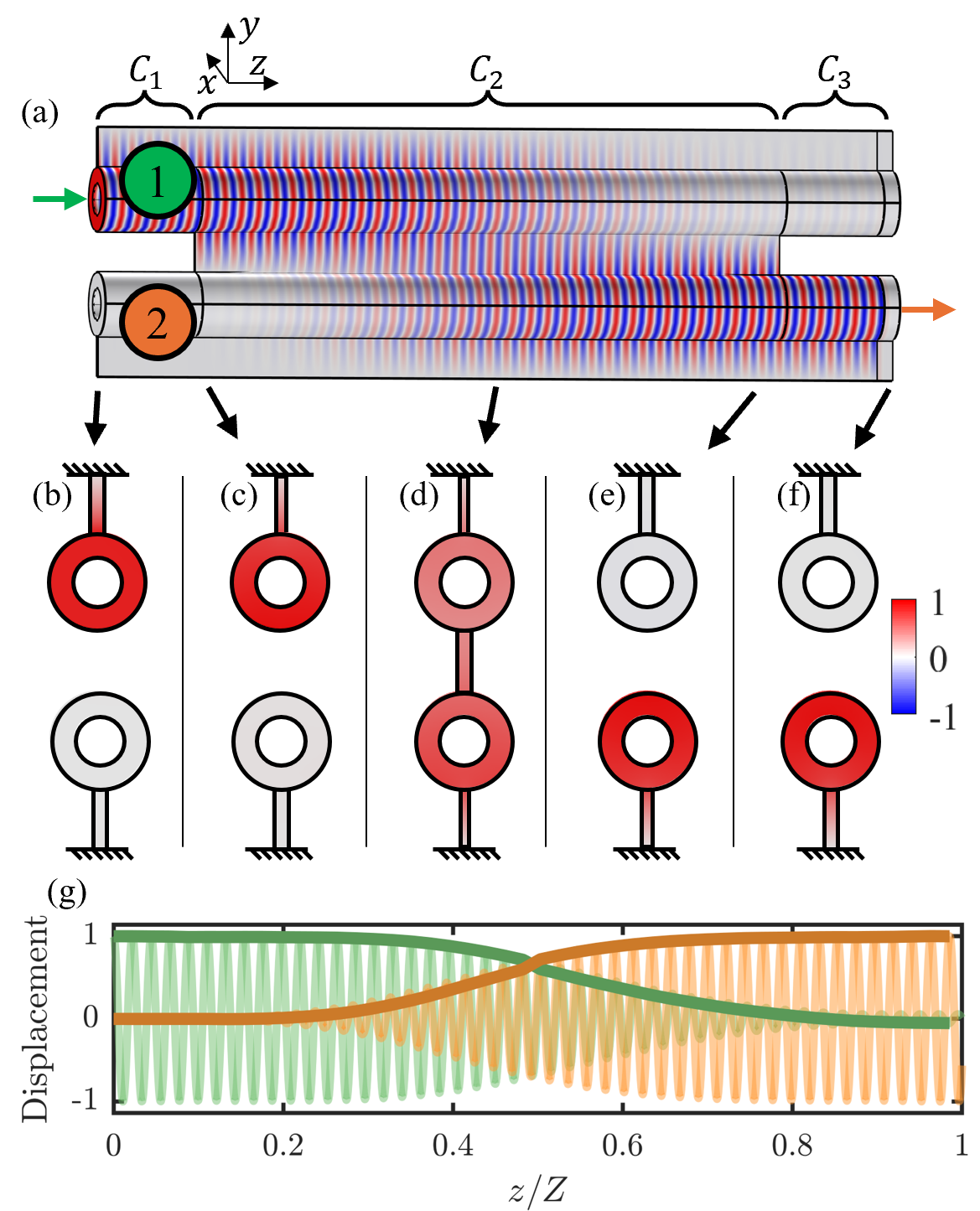}
    \caption{(a) A composite elastic waveguide consisting of two coupled rods. The waveguide is designed to exhibit the holonomy $\mc{H}_\mathrm{1,2D}$. The three stages of variations in cross-section are marked $C_1$, $C_2$, and $C_3$. Rod 1 is excited at $z=0$ (green arrow) according to its longitudinally polarized mode, and the ensuing steady-state longitudinal displacement response is shown by the colormap. The output response at $z=Z$ is localized on rod 2 (orange arrow). The displacements are normalized against the input excitation. For clarity in the visualization, the z-axis is compressed by a factor of 40. (b-f) Cross section of the waveguide at $z=0$, $z=Z/8$, $z=Z/2$, $z=7Z/8$, and $z=Z$, respectively. The colormap indicates the normalized longitudinal displacement after factoring out the dynamical phase. (g) Longitudinal displacements of rod 1 (green) and rod 2 (orange) measured on a line parallel to the axis of the corresponding cylinder and lying on its surface. The thick lines denote the components of the displacement after factoring out the dynamical phase.}
    \label{fig:two-rod-exchange}
\end{figure}

The thickness of the plates $h_{ij}(z)$ vary over a length $Z$ such that the initial and final thickness values are identical, $h_{ij}(0)=h_{ij}(Z)$. Geometrically, the thickness values $h_{ij}(z)$ trace a loop $L$ parameterized by $z$ in parameter space (Fig.~\ref{fig:design-schematic}c). The parameter space loops begin at a point corresponding to identical uncoupled rods (black dot in Fig.~\ref{fig:design-schematic}c). The thickness of all grounding plates are $h^{(0)}=0.6h_{\mathrm{max}}$, and there are no connecting plates, $h_{ij}(0)=0$ ($i \neq j$) (Fig.~\ref{fig:two-rod-exchange}b). In other words, the coupling matrix is a scalar multiple of the identity matrix: $\mc{K}(0)=h^{(0)}\mc{I}_N$. Consequently, this configuration admits guided modes localized on each rod, $\mathbf{W}_i(0)=\mathbf{e}_i$, leading to an initial orthonormal frame $\mc{W}(0)=\mc{I}_N$. All modes possess identical wavenumbers equaling $\kappa_0^2=\omega^2/c^2-Gh^{(0)}/EAl_c$, which indicates that this initial configuration supports an $N$-fold degeneracy.

The loop has three segments: $C_1$, from $z=0$ to $z=Z/8$; $C_2$, from $z=Z/8$ to $z=7Z/8$; and $C_3$, from $z=7Z/8$ to $Z$ (Fig.~\ref{fig:design-schematic}c). The thickness variation in each segment is described; for explicit formulas, see SI Appendix, Sec.~5C. During $C_1$, the thickness of the grounding plates change from $h^{(0)}$ to equally spaced values between $0.35 h_\mathrm{max}$ and $0.85 h_{\mathrm{max}}$, and no coupling plates are introduced (Fig.~\ref{fig:two-rod-exchange}b,c). The final thickness of the $i$th grounding plate is $h^{(1)}_i=0.35 h_\mathrm{max} + \frac{i-1}{N-1}0.5h_\mathrm{max}$; these thickness values are collected into the vector $\mathbf{h}^{(1)}$. In terms of the coupling matrix, it evolves from $\mc{K}(0)=h^{(0)}\mc{I}_N$ to $\mc{K}(Z/8)=\mathrm{diag}\, \mathbf{h}^{(1)}$. Since the coupling matrix remains diagonal, the guided modes remain localized on individual rods, $\mathbf{W}_i(z)=\mathbf{e}_i$. However, their wavenumbers split from $\kappa_0$ to $\kappa_{1,i}$, where $ \kappa_{1,i}^2=\omega^2/c^2-G h^{(1)}_i/EAl_c$.

Next, $C_2$ is a path that introduces coupling plates and changes the thickness of the plates in a nontrivial manner in order to acquire the holonomy $\mc{H}$ (Fig.~\ref{fig:design-schematic}c). The idea is that the thickness variation should preserve the wavenumbers while rotating the orthonormal frame of guided modes $\mc{W}(z)$ from the identity matrix $\mc{I}_N$ at $z=Z/8$ to the holonomy matrix $\mc{H}$ at $z=7Z/8$. This rotation is implemented through a matrix $\mc{R}(z)$, defined by smoothly increasing the angles in the Givens decomposition of $\mc{H}$ (Eq.~\ref{eqn:givens-holonomy}) from zero to their actual values (SI Appendix, Sec.~4A). Explicitly,
\begin{equation}
    \mc{R}(z)=\mc{G}(N-1,N,f(z)\theta_{N-1,N})\dots\mc{G}(1,2,f(z)\theta_{12})\;,
\end{equation}
where $f(z)=\sin^2 \left( \frac{\pi}{2} \frac{8z-Z}{6Z}\right)$ is a smooth function such that $f(Z/8)=0$ and $f(7Z/8)=1$. Clearly,
\begin{align}
    \label{eqn:rot-start}
    \mc{R}(Z/8)&=\mc{I_N}\;,\\
    \label{eqn:rot-end}
    \mc{R}(7Z/8)&=\mc{H}\;.
\end{align}
To effect this rotation of the orthonormal frame, the coupling matrix must evolve as
\begin{equation}
\label{eqn:coupling-matrix-c2}
    \mc{K}(z)=\mc{R}(z) (\mathrm{diag} \mathbf{h}^{(1)} ) \mc{R}(z)^T
\end{equation}
(see SI Appendix, Sec.~5C for a detailed derivation). From the coupling matrix, the required thickness variation $h_{ij}(z)$ is obtained by using Eq.~\ref{eqn:coupling-matrix}.

Physically, $C_2$ begins with a set of uncoupled rods whose grounding plates have nonidentical thickness (Fig.~\ref{fig:two-rod-exchange}c). Plate connections are then gradually introduced between the rods (Fig.~\ref{fig:two-rod-exchange}d), before being removed to return to a configuration of uncoupled rods (Fig.~\ref{fig:two-rod-exchange}e). The final configuration differs from the initial one by a permutation of the thickness of the grounding plates.

Finally, in $C_3$, the thickness of the grounding plates is returned to the initial value of $h^{(0)}$ without coupling the rods, closing the loop in parameter space (Fig.~\ref{fig:design-schematic}c). The coupling matrix remains diagonal and reaches $h^{(0)}\mc{I}_N$. Thus, the orthonormal frame of guided modes remains fixed at $\mc{H}$, and the wavenumbers return to $\kappa_0$.

\subsubsection{Waveguide length}
The length of the waveguide is chosen so that the rate of variation in thickness is slow enough to satisfy the adiabatic condition (SI Appendix, Sec.~1B) in $C_2$~\cite{bretherton_propagation_1968,berry_quantal_1984,budden_phase_1976}; the condition on the length is $Z \gg \kappa_0 r_2^2$ (derived in SI Appendix, Eq.~116). In practice, the discrete model is simulated for increasing values of $Z$, and a value resulting in satisfactory performance is selected (SI Appendix, Sec.~5D).   

We note that the values of $Z$ are large, leading to extremely slender waveguides: the ratio of waveguide length to nominal cross-sectional size, $Z/(r_2+l_c)$, exceeds 1000. Nevertheless, future design implementations could alleviate this issue by using alternate platforms such as coupled-resonator elastic waveguides~\cite{escalanteDispersionRelationCoupledresonator2013} or by leveraging shortcuts to adiabaticity~\cite{guery-odelinShortcutsAdiabaticityConcepts2019,song_shortcuts_2025}.

\section{Observing the holonomy}
\subsection{Exchanging modes in two coupled rods}
The above design principle is practically illustrated using an example of two coupled rods (Fig.~\ref{fig:two-rod-exchange}a). The waveguide is designed to exhibit a holonomy represented by the matrix 
\begin{equation}
\label{eqn:two-rods-holonomy-matrix}
    \mc{H}_{1,2\rm{D}}=\mc{G}(1,2,\pi/2)=
    \begin{pmatrix}
        0 & -1 \\ 
        1 & 0
    \end{pmatrix}
    \;.
\end{equation} 
The holonomy is numerically observed in Fig.~\ref{fig:two-rod-exchange}a by exciting one end of rod 1 with an axial force and measuring the steady-state response in frequency domain (Materials and Methods). In all numerical simulations, the operating frequency is fixed at $\omega_\mathrm{max}$. In addition, Fig.~\ref{fig:two-rod-exchange}g plots the longitudinal displacements in the two cylinders. The thick solid lines denote the envelope of the displacement profile for each cylindrical waveguide or, equivalently, the displacement amplitudes after factoring out the dynamical phase (Materials and Methods).

When the initial displacement $\hb{w}(0)=\mathbf{e}_1$ (Fig.~\ref{fig:two-rod-exchange}b) propagates through the section of the waveguide associated with the path $C_1$, its displacement profile remains unchanged after factoring out amplitude and phase factors (Fig.~\ref{fig:two-rod-exchange}c). In mathematical terms this means that $\hb{w}(Z/8)=\mathbf{e}_1 a(Z/8) e^{\ii \phi(Z/8)}$, where $a(Z/8)$ is the variation in amplitude and $\phi(Z/8)$ is the dynamical phase (see SI Appendix, Sec.~1A for a derivation).

Next, the wave $\hb{w}(z)$ propagates through the waveguide segment corresponding to the path $C_2$, during which its variation is dictated by adiabatic evolution~\cite{bretherton_propagation_1968,tromp_berry_1992,kumar_role_2024,budden_phase_1976}. Adiabatic evolution implies that the incident wave $\hb{w}(Z/8)$, initially proportional to the first guided mode $\mathbf{W}_1(Z/8)=\mathbf{e}_1$ of the cross section at $Z/8$, remains proportional to the corresponding first guided mode $\mathbf{W}_1(z)$ of the cross section at $z$ (see SI Appendix, Sec.~1B for a derivation). By the definition of parameter variation in $C_2$ (Eq.~\ref{eqn:coupling-matrix-c2}), the first guided mode at $z$ is $\mathbf{W}_1(z)=\mc{R}(z)\mathbf{e}_1$. Thus, the wave evolves as $\hb{w}(z)=\mc{R}(z) \mathbf{e}_1 a(z) e^{\ii \phi(z)}$, as illustrated in Figs.~\ref{fig:two-rod-exchange}c-e. The displacement is initially confined to rod 1 (Fig.~\ref{fig:two-rod-exchange}c), since
\begin{equation}
    \hb{w}(Z/8) \propto \mc{R}(Z/8)\mathbf{e}_1= \mc{I}_N \mathbf{e}_1=\mathbf{e}_1\;,
\end{equation}
where Eq.~\ref{eqn:rot-start} was used. As $z$ increases, it distributes across multiple rods (Fig.~\ref{fig:two-rod-exchange}d). At $z=7Z/8$, it localizes on rod 2 (Fig.~\ref{fig:two-rod-exchange}e), consistent with
\begin{equation}
    \hb{w}(7Z/8) \propto \mc{R}(7Z/8)\mathbf{e}_1= \mc{H}_{1,2\rm{D}}\mathbf{e}_1=\mathbf{e}_2\;,
\end{equation}
where Eq.~\ref{eqn:rot-end} was used.

Finally, as the wave propagates in the segment corresponding to thickness variation in $C_3$, its displacement remains localized in rod 2. At the end of the waveguide, the displacement is $\hb{w}(Z) \propto \mc{H}_{1,2\rm{D}}\mathbf{e}_1=\mathbf{e}_2$.  

In summary, the initial excitation on rod 1, corresponding to $\mathbf{e}_1=(1,0)^T$, evolves to the terminal response in rod 2 $\mathbf{e}_2=(0,1)^T$. A similar simulation, where rod 2 is excited, confirms that the initial displacement $\mathbf{e}_2$ evolves to $-\mathbf{e}_1$ (SI Appendix, Fig.~S4), in agreement with $\mc{H}_{1,2\rm{D}} \mathbf{e}_2 = -\mathbf{e}_1$. Together, these observations verify that the elastic waveguide achieves the desired holonomy matrix $\mc{H}_{1,2\rm{D}}$. 

\subsection{Exchanging modes in three coupled rods}
\begin{figure}
    \centering
    \includegraphics[width=\linewidth]{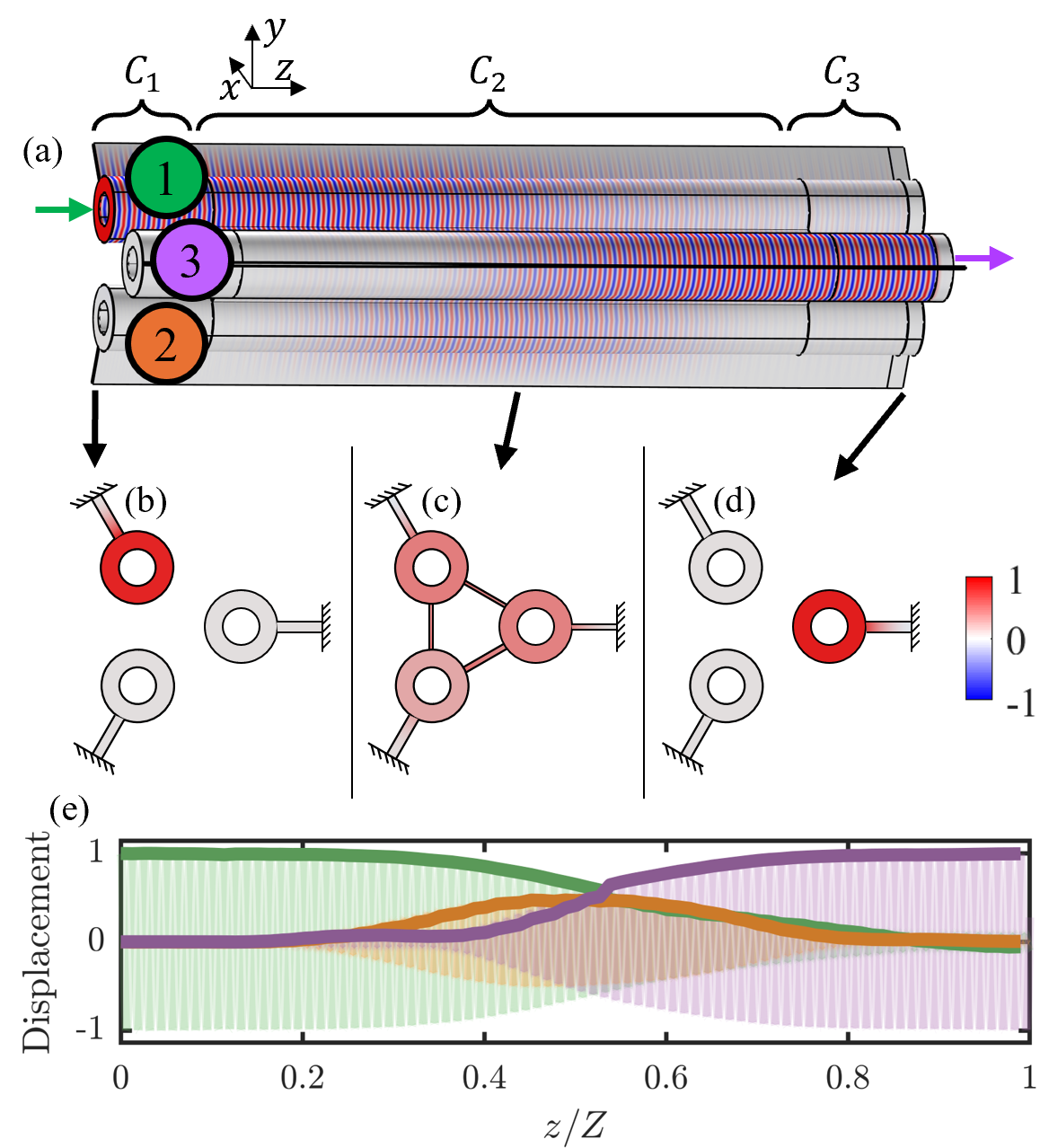}
    \caption{(a) The composite elastic waveguide WG2 comprising three coupled rods. The waveguide is designed to exhibit the holonomy $\mc{H}_2$. The three stages of variation of the cross-section are marked $C_1$, $C_2$, and $C_3$. Rod 1 is excited at $z=0$ (green arrow) according to its longitudinally polarized mode, and the ensuing steady-state longitudinal displacement response is shown by the colormap. The output response at $z=Z$ is localized on rod 3 (purple arrow). The displacements are normalized against the input excitation. For clarity in the visualization, the z-axis is compressed by a factor of 70. (b-d) Cross section of the waveguide at $z=0$, $z=Z/2$, and $z=Z$, respectively. The colormap indicates the normalized longitudinal displacement after factoring out the dynamical phase. (e) Longitudinal displacements of rod 1 (green), rod 2 (orange), and rod 3 (purple) measured on a line parallel to the axis of the corresponding cylinder and lying on its surface. The thick lines denote the components of the displacement after factoring out the dynamical phase.}
    \label{fig:three-rod-exchange}
\end{figure}

Using the same design principle, a waveguide comprising three coupled rods (Fig.~\ref{fig:three-rod-exchange}a) is constructed to achieve a holonomy corresponding to the matrix
\begin{equation}
\label{eqn:three-rods-holonomy-matrix}
    \mc{H}_2=\mc{G}(2,3,\pi/2)\mc{G}(1,2,\pi/2)=
    \begin{pmatrix}
        0 & -1 & 0\\
        0 & 0 & -1\\
        1 & 0 & 0
    \end{pmatrix}
    \;.
\end{equation}
We label this waveguide `WG2'.

Figure~\ref{fig:three-rod-exchange}a shows the steady-state response of the waveguide when it is excited at one end with an axial displacement on rod 1. The wave shifts from rod 1 to rod 3 as it reaches the end of the cross-section perturbation (Figs.~\ref{fig:three-rod-exchange}b-d). Figure~\ref{fig:three-rod-exchange}e further verifies that the longitudinal displacement, after having factored out the dynamical phase, evolves from $\mathbf{e}_1$ to $\mathbf{e}_3$, in accordance with $\mc{H}_2 \mathbf{e}_1 = \mathbf{e}_3$ (Eq.~\ref{eqn:three-rods-holonomy-matrix}). Similarly, an excitation in rod 2, $\mathbf{e}_2$, evolves to a wave in rod 1 with a sign-reversal, $\mc{H}_2\mathbf{e}_2=-\mathbf{e}_1$, and an excitation in rod 3, $\mathbf{e}_3$, evolves to a wave in rod 2 with a sign-reversal, $\mc{H}_2\mathbf{e}_3=-\mathbf{e}_2$ (SI Appendix, Fig.~S5).

\subsection{Verification of nonabelian behavior}
The nonabelian nature of the holonomy is explored in a system of three coupled rods. As a preliminary step, the two-rod system is extended to a three-rod system by adding an uncoupled rod. We label this waveguide `WG1'. The corresponding holonomy is 
\begin{equation}
\label{eqn:two-rods-holonomy-matrix-3D}
    \mc{H}_{1}=
    \begin{pmatrix}
        \mc{H}_{1,2\rm{D}} & 0 \\
        0 & 1
    \end{pmatrix}
    =
    \begin{pmatrix}
        0 & -1 & 0 \\ 
        1 & 0 & 0 \\
        0 & 0 & 1
    \end{pmatrix}
    \;.
\end{equation}
\begin{figure*}[!tb]
    \centering
    \includegraphics[width=\linewidth]{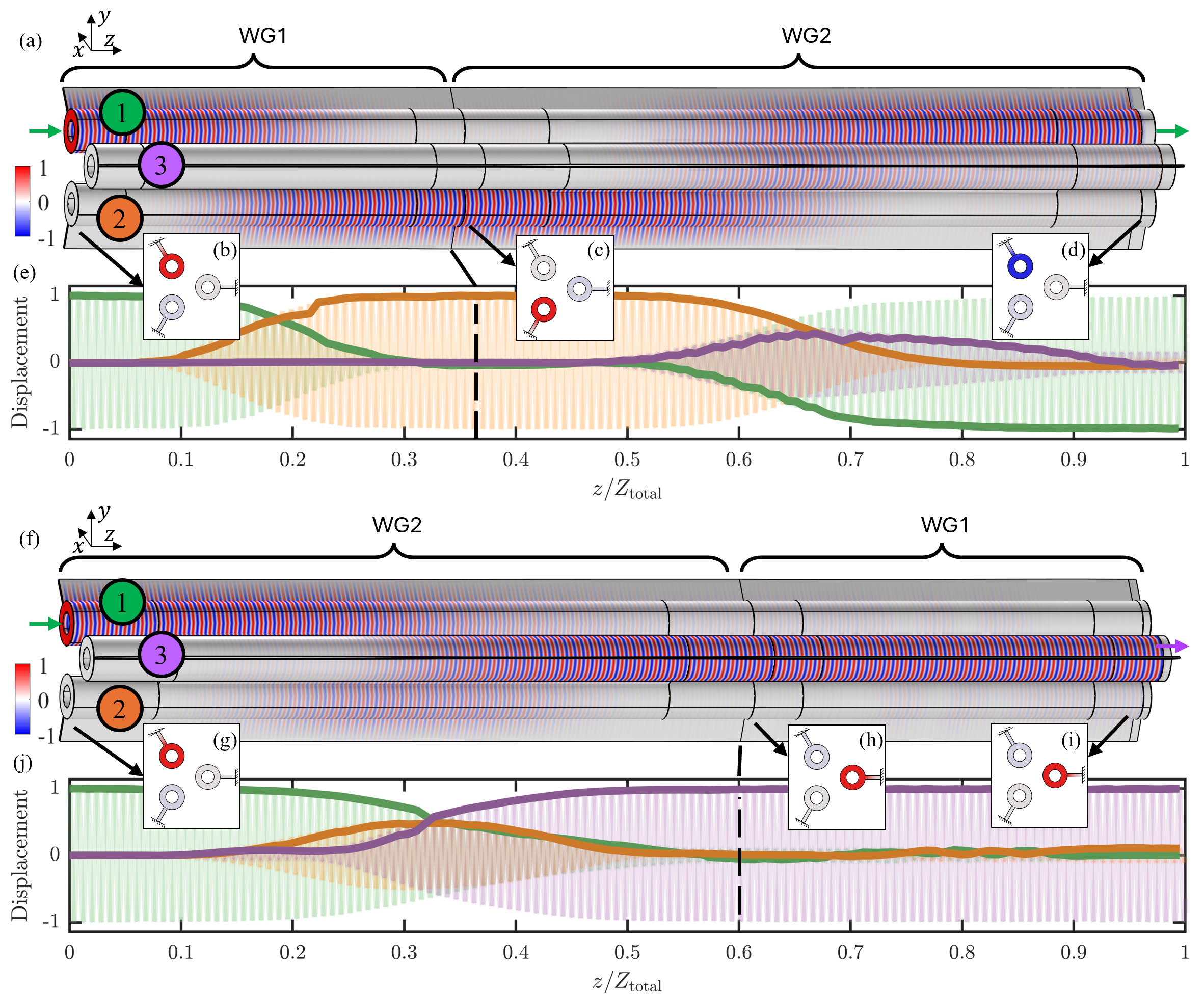}
    \caption{(a) The elastic waveguide WG12 consisting of waveguide WG1 followed by waveguide WG2. Rod 1 is excited at $z=0$ (green arrow, left end) according to its longitudinally polarized mode, and the ensuing steady-state longitudinal displacement response is shown by the colormap. The output response is localized on rod 1 (green arrow, right end). The displacements are normalized against the input excitation. For clarity in the visualization, the z-axis is compressed by a factor of 55. (b-d) Cross section of the waveguide at the start of WG1, at the end of WG1, and at the end of WG2, respectively. The colormap indicates the normalized longitudinal displacement after factoring out the dynamical phase. (e) Longitudinal displacements of rod 1 (green), rod 2 (orange), and rod 3 (purple) measured on a line parallel to the axis of the corresponding cylinder and lying on its surface. The thick lines denote the components of the displacement after factoring out the dynamical phase. (f) The elastic waveguide WG21 and its longitudinal displacement response when rod 1 is excited at $z=0$. For clarity in the visualization, the z-axis is compressed by a factor of 55. (g-i) Cross section of the waveguide at the start of WG2, at the end of WG2, and at the end of WG1, respectively. (j) Longitudinal displacements of the rod 1 (green), rod 2 (orange), and rod 3 (purple) measured on a line parallel to the axis of the corresponding cylinder and lying on its surface.}
    \label{fig:nonabelian-behavior}
\end{figure*}

To verify the nonabelian nature, waveguides are connected in different order and their holonomies are observed. The first waveguide, denoted `WG12', comprises WG1 followed by WG2 (Fig.~\ref{fig:nonabelian-behavior}a). Here, an excitation $\mathbf{e}_1$ on rod 1 (Fig.~\ref{fig:nonabelian-behavior}b) propagates to rod 2 at the end of WG1 (Fig.~\ref{fig:nonabelian-behavior}c), since $\mc{H}_1 \mathbf{e}_1 = \mathbf{e}_2$, and then transfers back to rod 1 at the end of WG12 (Fig.~\ref{fig:nonabelian-behavior}d), since $\mc{H}_2(\mc{H}_1 \mathbf{e}_1)=\mc{H}_2 \mathbf{e}_2=-\mathbf{e}_1$. The spatial evolution of the displacement is shown in Fig.~\ref{fig:nonabelian-behavior}e. By repeating this analysis for excitations on the second and third rods, $\mathbf{e}_2$ and $\mathbf{e}_3$, the holonomy of waveguide WG12 is
\begin{equation}
\label{eqn:holonomy-12}
    \mc{H}_{12}=\mc{H}_2\mc{H}_1=
    \begin{pmatrix}
        -1 & 0 & 0 \\ 
        0 & 0 & -1 \\
        0 & -1 & 0
    \end{pmatrix}
    \;.
\end{equation}

The second waveguide, denoted `WG21', exchanges the order of the constituent waveguides, so that the composite waveguide consists of WG2 followed by WG1 (Fig.~\ref{fig:nonabelian-behavior}f). In WG21, an excitation $\mathbf{e}_1$ on rod 1 (Fig.~\ref{fig:nonabelian-behavior}g) propagates to rod 3 at the end of WG2 (Fig.~\ref{fig:nonabelian-behavior}h), since $\mc{H}_2\mathbf{e}_1=\mathbf{e}_3$, and then stays on rod 3 at the end of WG21 (Fig.~\ref{fig:nonabelian-behavior}i), since $\mc{H}_1(\mc{H}_2\mathbf{e}_1)=\mc{H}_1\mathbf{e}_3=\mathbf{e}_3$. The spatial evolution of the displacements is shown in Fig.~\ref{fig:nonabelian-behavior}j. The holonomy matrix of WG21 is 
\begin{equation}
\label{eqn:holonomy-21}
    \mc{H}_{21}=\mc{H}_1\mc{H}_2=
    \begin{pmatrix}
        0 & 0 & 1 \\ 
        0 & -1 & 0 \\
        1 & 0 & 0
    \end{pmatrix}
    \;.
\end{equation}

Thus, the same excitation on rod 1, $\mathbf{e}_1$, propagates to rod 1 in WG12 and to rod 3 in WG21. More generally, the total holonomy of WG12 differs from that of WG21: $\mc{H}_{12} \neq \mc{H}_{21}$ (Eqs.~\ref{eqn:holonomy-12},~\ref{eqn:holonomy-21}). That is, exchanging the constituent waveguides changes the final holonomy, signifying its nonabelian nature. Thus, the dynamics of WG12 and WG21 physically realize a fundamental principle of nonabelian behavior: exchanging the sequence of operations changes the observed response.

\subsection{Broadband performance and robustness}
The nonabelian behavior of the waveguides extends across the range of frequencies for which the adiabatic condition is satisfied (SI Appendix, Eq.~17) and the longitudinal modes are isolated from torsional and flexural modes. While the present simulations used operating frequencies of $\omega_\mathrm{max}$, the nonabelian behavior is verified to persist from $0.67 \omega_\mathrm{max}$ to $1.25 \omega_\mathrm{max}$ (SI Appendix, Fig.~S6).

The nonabelian dynamics are also robust against several perturbations to the waveguide's geometry, or equivalently, to perturbations in the loop $L$ traversed in parameter space. Firstly, arbitrary perturbations to the plates or the rods in the $C_2$ segment of the waveguides do not change the holonomy, provided that the perturbations maintain the separation between wavenumbers to satisfy the adiabatic condition (SI Appendix, Eq.~17). Secondly, arbitrary perturbations to the grounding plates or the rods in the $C_1$ segment of the waveguides do not change the holonomy. These perturbations are visualized in parameter space as deforming loop $L$ into $L'$ (Fig.~\ref{fig:design-schematic}c). This robustness is verified in SI Appendix, Fig.~S8, where the nonabelian behavior observed in Fig.~\ref{fig:nonabelian-behavior} persists despite perturbations to the plate thickness and cylinder inner radii of up to $0.1 h^{(0)}$ and $0.1r_1$, respectively (see SI Appendix, Fig.~S7 for the perturbed geometrical parameters).

Finally, longer waveguides exhibit the same holonomy. More generally, changes in the rate of thickness variation leave the holonomy invariant as long as the adiabatic condition is satisfied (SI Appendix, Eq.~17). In other words, the holonomy primarily depends on the geometry of the loop $L$ in the parameter space of thicknesses, not on its parametrization. 

On the one hand, these properties imply that nonabelian holonomy of an elastic waveguide tolerates manufacturing defects and operational wear. On the other hand, this robustness enables straightforward design of continuous elastic waveguides using an approximate discrete model (Eq.~\ref{eqn:rod-full-eom}) without any fine-tuning. As a result, the present design strategy provides a simple method to realize robust nonabelian holonomies in elastic waveguides. 

\section{Theoretical analysis and generalizations}

\subsection{Characterizing achievable nonabelian holonomies}
Beyond specific realizations, the entire set of nonabelian holonomies that can be realized by the above design principle is now investigated. This set of nonabelian holonomies forms the holonomy group $H^N_1$, which is shown to be related to the well-known braid groups.

\subsubsection{Holonomy group}
Recall that the proposed design procedure can, in general, create waveguides that exhibit holonomies expressed as products of Givens matrices (Eq.~\ref{eqn:givens-holonomy}). Moreover, any orientation-preserving signed permutation matrix can be written as a product of Givens matrices (see SI Appendix, Sec.~4C). It follows that the holonomy group $H^N_1$ coincides with the group of orientation-preserving signed permutation matrices. Thus, the holonomy group $H^N_1$ and the matrices corresponding to its elements may be presented as
\begin{equation}
\begin{aligned}
\label{eqn:nondegen-holonomy-group}
H^N_1 &=
\left\{
  \mc{P}\, \mathrm{diag}(a_1,\dots,a_N) \;\middle|\;
  \begin{aligned}
  & \mc{P}\in \mathfrak{S}_N, \; a_i = \pm 1, \\
  & \left( \textstyle{\prod_i} a_i \right) (\det \mc{P})=1
  \end{aligned}
\right\}\\
& \cong (\bb{Z}_2^N \rtimes \mathfrak{S}_N) \cap SO(N) \\
&\cong  \bb{Z}_2^{N-1} \rtimes \mathfrak{S}_N\;,
\end{aligned}
\end{equation}
where $\mc{P}$ is a permutation matrix of size $N$, $\mathfrak{S}_N$ is the symmetric group of degree $N$, $SO(N)$ is the special orthogonal group in dimension $N$, $\bb{Z}_2$ is the cyclic group of order 2, and $\rtimes$ denotes the semi-direct product. The group $H^N_1$ is nonabelian (for $N\geq3$).

$H^N_1$ enlarges the holonomy group, $H^N_{1,\mathrm{GP}}$, achieved by typical geometric phases in nondegenerate systems. Holonomies of $H^N_{1,\mathrm{GP}}$ are limited to changing the signs of guided modes while preserving the overall orientation~\cite{berry_quantal_1984,kumar_role_2024}. That is, $H^N_{1,\mathrm{GP}}\cong  \bb{Z}_2^{N-1}$, which is an abelian group. By allowing loops in parameter space to cross degeneracies, the preceding analysis shows that $H^N_{1,\mathrm{GP}}$ is augmented by permutations in $\mathfrak{S}_N$ to yield the nonabelian holonomy group $H^N_1$. The role of crossing degeneracies is mathematically formalized in terms of fiber bundles and connections in SI Appendix, Sec.~3.

\subsubsection{Connection to braid groups}

Notice that all holonomy matrices in $H^N_1$ can be described by a sequence of signed permutations that exchange guided modes localized on the $i$th and $(i+1)$th rods, where $i=1,\dots,N-1$. That is, $H^N_1$ is generated by the matrices $\mc{H}^g_i=\mc{G}(i,i+1,\pi/2)$. Furthermore, these generating matrices satisfy
\begin{equation}
\label{eqn:holonomy-braid-relations}
    \mc{H}^g_i \mc{H}^g_{i+1}\mc{H}^g_i=\mc{H}^g_{i+1}\mc{H}^g_i\mc{H}^g_{i+1}\;.
\end{equation}

These simple relations connect the holonomy group $H^N_1$ to the braid group on $N$ strands, $B_N$. An element of a braid group $B_N$ may be visualized as a braid created with $N$ vertical strands~\cite{nayak_non-abelian_2008}. Any braid can be created by sequentially exchanging adjacent $i$th and $(i+1)$th strands using the element $\tau_i$, where $i=1,\dots,N-1$. Further, these generating braids satisfy $\tau_i \tau_{i+1} \tau_i = \tau_{i+1} \tau_i \tau_{i+1}$, analogous to Eq.~\ref{eqn:holonomy-braid-relations}. 

Formally, identifying the braid $\tau_i$ with the matrix $\mc{H}^g_i$ defines a representation of the braid group $B_N$ that factors through the holonomy group $H^N_1$. These results link the nonabelian behavior achieved by the elastic waveguides to the intuitive notion of braiding. Furthermore, representations of the braid group form a critical component of topological quantum computation and its classical emulations~\cite{nayak_non-abelian_2008,barlas_topological_2020,zou_experimental_2023,ezawa_non-abelian_2020}. By exploiting this connection, holonomies of $H^N_1$ may provide an avenue towards elastic-wave-based physical computation. 

\subsection{Systems evolving on degenerate manifolds}
Thus far, parameter variations crossing $N$-fold degeneracies in otherwise nondegenerate systems expanded the achievable holonomies from geometric-phase-based sign changes to signed permutations. The underlying theory naturally extends to coupled systems that maintain degeneracies. The main results are summarized here; see SI Appendix, Sec.~1D.2 and Sec.~3B for derivations.

Consider an idealized coupled resonator system with $N$ degrees of freedom governed by Eq.~\ref{eqn:eom}. At any $z$, the corresponding dynamical matrix is assumed to possess $n$ distinct eigenvalues $\{k_1(z)^2, \dots, k_n(z)^2\}$, where each eigenvalue is $m$-fold degenerate and is associated with an $m$-dimensional eigenspace. 

Loops in parameter space preserving eigenvalue degeneracies result in holonomies that are matrix-valued Wilczek-Zee phases~\cite{wilczek_appearance_1984}. Such a holonomy performs ``internal rotations'' within eigenspaces while preserving the overall orientation. That is, a system initialized in an eigenvector in the $i$th eigenspace, after a loop in parameter space, evolves to another eigenvector also within the $i$th eigenspace. The resulting holonomy group is $H^n_{m,\mathrm{WZ}}\cong (O(m))^n \cap SO(mn)$, where $O(m)$ is the orthongonal group in $m$ dimensions.

Next, similar to earlier sections, consider systems evolving along loops in parameter space that begin and end at an $N$-fold degeneracy, while maintaining $m$-fold degeneracies at all other points. The resulting holonomy performs an external permutation of the eigenspaces in addition to internal rotations. That is, a system initialized in an eigenvector in the $i$th eigenspace, after a loop in parameter space, evolves to an eigenvector in the $j$th eigenspace, where $i \neq j$ generically. The holonomy group is
\begin{equation}
\label{eqn:real-degen-holonomy}
    H^n_{m} \cong (O(m)^n \rtimes \mathfrak{S}_n) \cap SO(mn)\;,
\end{equation}
which contains the Wilczek-Zee holonomy group $H^n_{m,\mathrm{WZ}}$ as a subgroup. When $n=N$ and $m=1$, the eigenvalues are generically nondegenerate, and one recovers the holonomy group in Eq.~\ref{eqn:nondegen-holonomy-group}, noting that $O(1)\cong \bb{Z}_2$. 

These holonomy matrices can be used to obtain an infinite representation of the braid group (SI Appendix, Sec.~1D.2), further expanding the potential of classical emulations for topological quantum computation. While its practical implementation is certainly challenging in elastic systems due to the requirement of maintaining $m$-fold degeneracies, this strategy could be explored using existing acoustic~\cite{chen_classical_2022} or photonic~\cite{zhang_non-abelian_2022} waveguides.

\subsection{First-order systems}
The above results on holonomies of coupled rods or resonators, which are governed by real-valued second-order differential equations (Eq.~\ref{eqn:eom}), extend to systems governed by complex-valued first-order differential equations (SI Appendix, Sec.~2), including time-varying quantum systems governed by the Schr\"odinger equations as well as spatially varying photonic, acoustic, and elastic waveguides modeled using the coupled mode approximation~\cite{huang_coupled-mode_1994}. 

For such systems, parameter variations along a loop crossing a degeneracy results in a holonomy that performs rotations within an eigenspace, permutes eigenspaces, and preserves the orientation, similar to real-valued systems. However, rotations now imply $m$-dimensional unitary transformations in $U(m)$. Thus, for a generically nondegenerate system, the usual Berry phase acquired by an eigenvector is augmented with permutations, and the resulting holonomy group is $K^N_1 \cong (U(1)^N \rtimes \mathfrak{S}_N) \cap SU(N)$ (SI Appendix, Sec.~2D.1), in analogy with Eq.~\ref{eqn:nondegen-holonomy-group}. Similarly, for systems that maintain $m$-fold degeneracies for each of its $n$ distinct eigenvalues during parameter variation, the typical Wilczek-Zee phase is augmented by block permutations, and the resulting holonomy group is $K^n_{m}\cong  (U(m)^n \rtimes \mathfrak{S}_n) \cap SU(mn)$ (SI Appendix, Sec.~2D.2), analogous to Eq.~\ref{eqn:real-degen-holonomy}.

\section{Discussion}
The existing literature has abundantly shown that for a typical, slow, cyclic variation of the parameters in spatially varying waveguides, an excitation in the $i$th eigenspace propagates to an output response confined to the same eigenspace~\cite{berry_quantal_1984,bretherton_propagation_1968} if the parameter variation avoids $N$-fold degeneracies. In contrast, the main theoretical result derived and exploited in this study is that cyclic parameter variations that cross $N$-fold degeneracies can propagate an initial excitation in the $i$th eigenspace to a final response in a generally different $j$th eigenspace. Therefore, crossing the $N$-fold degeneracy in parameter space enriches the conventional holonomies--geometric phases in nondegenerate systems and Wilczek-Zee phases in degenerate systems--by introducing eigenspace permutations, as summarized in Tab.~\ref{tab:holonomy_classification}.
\begin{table*}
\centering
\caption{Different types of systems and parameter variations, the resulting holonomy groups, and their properties. The system has $N$ degrees of freedom. For nondegenerate systems, there are $N$ distinct eigenvalues. For systems maintaining degeneracies, there are $n$ distinct eigenvalues, each with multiplicity $m>1$. The practical realization corresponding to a row is termed fine-tuned if parameter variations must maintain $m$-fold degeneracies and robust if parameter variations traverse nondegenerate systems.}
\footnotesize
\label{tab:holonomy_classification}
\begin{tabular}{cccccc}
\hline
System & Eigenvalues & \makecell[c]{Parameter variation crosses\\$N$-fold degeneracy} & Holonomy group & Type & Practical realization \\
\hline
\multirow{4}{*}{\begin{tabular}{c}
2nd order,\\
real-valued
\end{tabular}}
& \multirow{2}{*}{Nondegenerate}
& No  & Topological geometric phase, $\bb{Z}_2^{N-1}$    & Abelian    & Robust  \\
& 
& Yes & $\bb{Z}_2^{N-1} \rtimes \mathfrak{S}_N$ & Nonabelian & Robust  \\
\cline{2-6}
& \multirow{2}{*}{$m$-fold degenerate}
& No  & Wilczek-Zee phase, $O(m)^n \cap SO(mn)$    & Nonabelian    & Fine-tuned \\
& 
& Yes & $(O(m)^n \rtimes \mathfrak{S}_n) \cap SO(mn)$ & Nonabelian & Fine-tuned \\
\hline
\multirow{4}{*}{\begin{tabular}{c}
1st order,\\
complex-valued
\end{tabular}}
& \multirow{2}{*}{Nondegenerate}
& No  &  Berry phase, $U(1)^N  \cap SU(N)$   & Abelian    & Robust  \\
& 
& Yes & $(U(1)^N \rtimes \mathfrak{S}_N) \cap SU(N)$ & Nonabelian & Robust  \\
\cline{2-6}
& \multirow{2}{*}{$m$-fold degenerate}
& No  & Wilczek-Zee phase, $U(m)^n \cap SU(mn)$  & Nonabelian & Fine-tuned \\
& 
& Yes & $(U(m)^n \rtimes \mathfrak{S}_n) \cap SU(mn)$ & Nonabelian & Fine-tuned \\
\hline
\end{tabular}
\end{table*}

Within this framework, the topological geometric phase of elastic waveguides~\cite{kumar_role_2024} expands to the holonomy group $H^N_1$, which consists of signed permutations. Holonomies in $H^N_1$ are nonabelian and closely related to the braid group $B_N$. Furthermore, they are easier to implement than Wilczek-Zee phases because they do not require degeneracies to be maintained throughout the entire parameter loop. 

Parameter variations that generate holonomies in $H^N_1$ led to the main practical result of this study, that is, a robust strategy to design nonabelian continuous elastic waveguides. Given a signed permutation matrix $\mc{H}$, a composite waveguide comprising $N$ cylindrical rods coupled by thin plates was designed to implement the target holonomy $\mc{H}$. In this waveguide, an input excitation provided to the $i$th rod propagates to the $j$th rod, as dictated by $\mc{H}\mathbf{e}_i=\pm\mathbf{e}_j$. When two or more such waveguides are concatenated, nonabelian dynamics emerges so that the output response to the same longitudinal input excitation depends on the ordering of the waveguides. This nonabelian behavior persists over a range of operating frequencies and under perturbations to the waveguide geometry, highlighting the broadband and robust nature of the design principle.

These findings provide a concrete platform to leverage unique features of nonabelian physics in the design of continuous elastic metamaterials. The elastic realizations of holonomies, their relation to the braid groups, and concepts from holonomic computation point toward a clear pathway towards elastic-wave-based physical computation. Incorporating active elements could further enable reconfigurable nonabelian behavior, in which the order of constituent waveguide segments can be changed in situ to redirect propagating waves. In conclusion, nonabelian holonomies achieved by crossing degeneracies offer a glimpse into the realm of nonabelian elastic metamaterials and their potential technological applications.   

\appendix

\section{Givens rotation}
The $N$-dimensional Givens rotation $\mc{G}(i,j,\theta)$ rotates the plane spanned by $\mathbf{e}_i$ and $\mathbf{e}_j$ by an angle of $\theta$. The corresponding matrix is
\begin{equation}
\label{eqn:givens-matrix}
\mc{G}(i,j,\theta)=
\begin{pNiceMatrix}[last-row,last-col,margin]
1 & \cdots & 0 & \cdots & 0 & \cdots & 0 &  \\
\vdots & \ddots & \vdots &        & \vdots &        & \vdots &   \\
0 & \cdots & \cos \theta & \cdots & -\sin \theta & \cdots & 0 & i \\
\vdots &        & \vdots & \ddots & \vdots &        & \vdots &   \\
0 & \cdots & \sin \theta & \cdots & \cos \theta & \cdots & 0 & j \\
\vdots &        & \vdots &        & \vdots & \ddots & \vdots &   \\
0 & \cdots & 0 & \cdots & 0 & \cdots & 1 &   \\
 &  & i &  & j &  & 
\end{pNiceMatrix}
\;,
\end{equation}
where the entries outside the matrix label the corresponding row and column.

\section{Numerical simulations of wave propagation}
The numerical simulations of the elastic waveguides were performed using the solid mechanics module of the commercial finite element software COMSOL Multiphysics. The steady-state dynamics was obtained by computing the response of the system in the frequency domain. Rods and grounding plates were modeled as three-dimensional elastic elements, while the coupling plates were modeled using two-dimensional plate elements with varying thickness. On the forced end of the waveguide, the cross section of one rod was subject to a prescribed sinusoidal displacement along the longitudinal $z$ direction. The other end of the waveguide was terminated with a perfectly matched layer to minimize unwanted reflections. The model was meshed using a swept mesh with a maximum element size of 0.04167 m. The mesh size was chosen to guarantee at least 10 elements per wavelength.

\section{Factoring out the dynamical phase}
From the steady state simulations, the dynamical phase was factored out to facilitate the observation of the evolution of the instantaneous guided modes. From the $j$th rod, the longitudinal displacements $\Hat{w}_j(z)$ were extracted from a line parallel to its axis and lying on its surface. The exact position of the line is inconsequential because the displacement profiles associated with the longitudinal modes considered in the analysis were nearly constant across the cross-section. The displacements were collected into the vector $\hb{w}(z)$. When the $i$th rod is excited, the displacement field is approximately described by $\hb{w}(z)=a_i(z) \mathbf{W}_i(z) \cos(-\int_0^z k_i(z') \dd z')$, where $\mathbf{W}_i(z)$ is the $i$th guided mode associated with the cross section at $z$ (SI Appendix, Sec.~1). To factor out the dynamical phase term, $\cos(-\int_0^z k_i(z') \dd z')$, only the peaks of $\hb{w}(z)$ are considered. The locations of peaks in $\hb{w}(z)$ are in turn extracted by considering the peaks of the largest component of $\hb{w}(z)$. 

\begin{acknowledgments}
    M.K. and F.S. gratefully acknowledge the financial support of the National Science Foundation under grant \#2330957. R.K. acknowledges financial support from the Simons Foundation.
\end{acknowledgments}

\bibliography{references}

\end{document}